\documentclass[preprintnumbers,amsmath,amssymb,showpacs]{revtex4}

\usepackage{graphicx}
\usepackage{dcolumn}
\usepackage{bm}
\usepackage{booktabs}
\usepackage{CJK}
\usepackage{subfigure}

\begin{document}

\title{Scalable symmetry detector and its applications by using beam splitters and weak nonlinearities}

\author{Yingqiu He$^{1}$}
\author{Dong Ding$^{2}$}
\email{dingdong@ncist.edu.cn}
\author{Fengli Yan$^3$ }
 \email{flyan@hebtu.edu.cn}
\author{ Ting Gao$^4$ }
 \email{gaoting@hebtu.edu.cn}
\affiliation {$^1$ Department of Biomedical Engineering, Chengde Medical University, Chengde 067000, China\\
$^2$ Department of Basic Curriculum, North China Institute of Science and Technology, Beijing 101601, China\\
$^3$ College of Physics Science and Information Engineering, Hebei Normal University, Shijiazhuang 050024, China \\
$^4$ College of Mathematics and Information Science, Hebei Normal University, Shijiazhuang 050024, China}
\date{\today}

\begin{abstract}

We describe a method to detect twin-beam multiphoton entanglement based on a beam splitter and weak nonlinearities.
For the twin-beam four-photon entanglement, we explore a symmetry detector.
It works not only for collecting two-pair entangled states directly from the spontaneous parametric down-conversion process, but also for purifying them by cascading these symmetry detectors.
Surprisingly, by calculating the iterative coefficient and the success probability we show that with a few iterations the desired two-pair can be obtained from   a class of four-photon entangled states.
We then generalize the symmetry detector to $n$-pair emissions and show that it is capable of determining the number of the pairs emitted indistinguishably from the spontaneous parametric down-conversion source, which may contribute to explore multipair entanglement with a large number of photons.

\end{abstract}

\maketitle

\section{Introduction}

Since optical quantum systems provide some natural advantages, they are prominent candidates for quantum information processing \cite{NC2000}.
As a fundamental physical resource, the multiphoton entanglement plays a crucial role in optical quantum computing \cite{KLM2001, Kok2007, Pan2012}.
A standard entangled photon pair is created by means of the nonlinear optical process of spontaneous parametric down-conversion (SPDC) \cite{SPDC1970}.
In SPDC process, one may create photons entangled in various degrees of freedom, for example, polarization entanglement \cite{PDC1995,PDC1992BMM, PDC1993KSC}, path entanglement \cite{PDC1986HZ, PDC1990RT}, etc.

For creation of polarization-entangled photons, a simplified Hamiltonian  \cite{H2000KB, H2000SWZ} of the nonlinear interaction is given by $H_{\mathrm{SPDC}} = \text{i}\kappa(\hat{a}^{\dag}_{H}\hat{b}^{\dag}_{V}-\hat{a}^{\dag}_{V}\hat{b}^{\dag}_{H})+\mathrm{H.c.}$,
where $\hat{a}^{\dag}_{x}$ and $\hat{b}^{\dag}_{x}$ (with $x=H, V$) are respectively the creation operators with horizontal ($H$) or vertical ($V$) polarization in the spatial modes $a$ and $b$,
and $\kappa$ is a real-valued coupling constant depended on the nonlinearity of the crystal and the intensity of the pump pulse.
In the number state representation, the resulting photon state reads  \cite{H2000SWZ, LHB2001, SB2003}
\begin{equation}\label{psi}
  \left| \Psi\right\rangle= \frac{1}{\mathrm{cosh}^2\tau}\sum_{n=0}^{\infty}\sqrt{n+1}\mathrm{tanh}^n\tau|\psi_n^{-}\rangle,
\end{equation}
\begin{equation}\label{psi n}
|\psi_n^{-}\rangle= \frac{1}{\sqrt{n+1}} \sum_{m=0}^{n}(-1)^m |n-m\rangle_{a_{H}}|m\rangle_{a_{V}}|m\rangle_{b_{H}}|n-m\rangle_{b_{V}},
\end{equation}
where e.g. $|m\rangle_{a_{V}}$ means $m$ vertically polarized photons in spatial mode $a$, and $\tau=\kappa t/\hbar$ is the interaction parameter with $t$ being interaction time.
Each $|\psi_n^{-}\rangle$ represents the state of $n$ indistinguishable photon pairs with $\langle n\rangle=2\mathrm{sinh}^2\tau$.
It should be noted that $|\psi_n^{-}\rangle$ is different from the general multiphoton entangled state in which each photon represents a qubit. State (\ref{psi n}) is usually called the twin-beam multiphoton entangled state.

To avoid multipair emission events, in general,  $\tau$ is restricted to small enough, such that mainly the first-order term has been taken into account. For the higher-order terms, these twin-beam multiphoton entangled states have  interesting features \cite{LHB2001}, i.e. they are not only entangled in photon number for the spatial modes $a$ and $b$, but also entangled maximally in polarization degree of freedom. Unfortunately, up to now there are only a few reports  \cite{ORW1999, DD2001, WZ2001, JMO2004, NOOST2007Science} to  exploring these analog of a singlet state of two spin-${n/2}$ particles.

In this paper, we first focus on the twin-beam four-photon entangled states and design a quantum circuit of symmetry detector to evolve them by using a beam splitter (BS) and weak nonlinearities. By cascading symmetry detectors, we then propose a scheme of purifying the twin-beam two-pair entangled state in a near deterministic way. Finally, we generalize the present symmetry detector to high-order emissions.

\section{Symmetry detector based on beam splitter and weak nonlinearities}

\begin{figure}
  \includegraphics[width=3.2in]{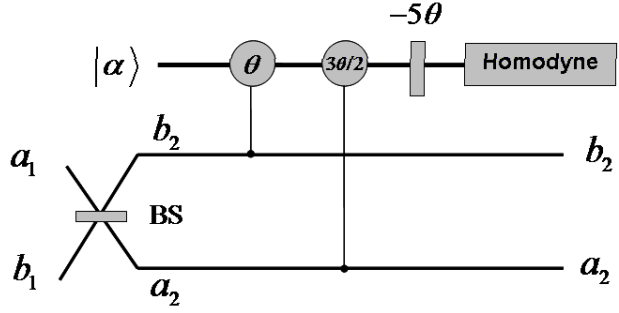}\\
  \caption{The schematic diagram of symmetry detector based on a beam splitter (BS) and weak nonlinearities. $a_{1}, b_{1}$ are input ports of a 50:50 BS, and $a_{2}, b_{2}$ are the corresponding outputs, respectively. $|\alpha\rangle$ is a coherent state in probe mode. $\theta$ and $3\theta/2$ are phase shifts on the coherent probe beam due to the interaction between photons in signal and probe modes. $-5\theta$ is a single phase gate.}
  \label{}
\end{figure}

Throughout the text, for simplicity we write $|m,n;r,s\rangle$ as an abbreviation for state $|m\rangle_{a_{H}}\otimes|n\rangle_{a_{V}}\otimes|r\rangle_{b_{H}}\otimes|s\rangle_{b_{V}}$
which means that there are $m$ horizontally polarized photons and $n$ vertically polarized photons in spatial mode $a$ and also there are $r$ horizontally and $s$ vertically polarized photons in spatial mode $b$.

We first restrict our attention to the four-photon entanglement and describe a method to explore symmetry detector for the twin-beam entangled states.
In general, consider a class of four-photon entangled states
\begin{equation}\label{4-photon-state-c}
  \left| \Phi\right\rangle= N(|2,0;0,2\rangle+|0,2;2,0\rangle - c|1,1;1,1\rangle),
\end{equation}
where $c$ is a constant and normalization factor $N$ satisfies $N^2 =1/(2+|c|^2)$. Without loss of generality, we may suppose coefficient $c$ to be real.

Consider a lossless $50:50$ BS with Hamiltonian
$H_{\mathrm{BS}} = -\text{i}\pi (\hat{a}^{\dag}_{}\hat{b}_{}- \hat{a}_{}\hat{b}^{\dag}_{})/4$, where $\hat{a}^{\dag}_{}$ ($\hat{b}^{\dag}_{}$ ) and $\hat{a}_{}$ ($\hat{b}_{}$ ) are respectively  creation and annihilation operators in the input spatial mode $a$ ($b$).
As shown in Fig.1, since the interference effect of BS, the input twin-beam state evolves
\begin{equation}\label{}
  \left| \Phi_{\mathrm{BS}}\right\rangle=N'[(1-c)(|2,2;0,0\rangle+|0,0;2,2\rangle)+(1+c)(|2,0;0,2\rangle+|0,2;2,0\rangle)-2|1,1;1,1\rangle],
\end{equation}
where $(N')^2=1/(4c^2+8)$.
An interesting consequence of this evolution is that the input state yields two possible cases, i.e., symmetric state (two photons are in one spatial mode and others are in another spatial mode) and alternatively asymmetric state (the output four photons are in the same spatial mode).

In order to distinguish between the symmetric state and the asymmetric state, we here consider quantum nondemolition detection  \cite{Imoto1985, MNS2005, MNBS2005, Barrett2005, LinHeBR2009, DY-PLA2013, HDYG2015} by using weak nonlinearities.
As an important nonlinear component for all-optical quantum computing, Kerr medium   \cite{SI1996, LI2000} is capable of evolving photons in signal and probe modes with the interaction Hamiltonian $H_{\mathrm{Kerr}} = \hbar \chi \hat{n}_{s}\hat{n}_{p}$, where $\chi$ is the coupling strength of the nonlinearity and $\hat{n}_{s}$ ($\hat{n}_{p}$) represents the number operator for the signal (probe) mode.
As a result, if there are $n$ photons in the signal mode, then it yields $n\theta$ in the probe mode, where $\theta  = \chi t$ is a phase shift on the coherent probe beam induced by the interaction via Kerr media and $t$ represents the interaction time.
Since the Kerr nonlinearities are extremely weak  \cite{LI2001, Kok2008}, we here only take small but available phase shifts into account.

As shown in Fig.1, after an overall interaction between the photons with Kerr media, the combined system $\left| \Phi_{\mathrm{BS}}\right\rangle\otimes|\alpha\rangle$ then evolves as
\begin{equation}\label{}
  \left| \Phi_{\mathrm{CK}}\right\rangle
  =\sqrt{1-P_{}}(|2,2;0,0\rangle |\alpha\text{e}^{\text{i}\theta}\rangle +|0,0;2,2\rangle |\alpha\text{e}^{-\text{i}\theta}\rangle)/\sqrt{2}
  + \sqrt{P_{}}N_{1}(|2,0;0,2\rangle+|0,2;2,0\rangle - c_{1}|1,1;1,1\rangle)|\alpha\rangle,
\end{equation}
where $c_{1}=2/(1+c)$ is the derived coefficient connected with the original $c$, $P_{}=1/\{1+(1-c)^2/[2+(1+c)^2]\}$ and $N_{1}^2=1/(2+c_{1}^2)$ are respectively the probability and normalization factor for the symmetric state.

We next turn to the question of how to project the signal photons into the symmetric state or the asymmetric state.
For a real coherent state, generally, one may perform an $X$ homodyne measurement   \cite{NM2004, DYG2014, HDYG2015OE} with the quadrature operator $\hat{x}=\hat{a}+\hat{a}^\dag$.
In terms of the result  \cite{Quantum-Noise2000} $\langle x|\alpha\rangle=({2\pi })^{ - 1/4}\text{exp}[-(\text{Im}(\alpha))^2-(x-2\alpha)^2/4]$,
after the $X$ homodyne measurement on the probe beam, for $x> \alpha(1+\text{cos}\theta)$, one can obtain the symmetric state
\begin{equation}\label{}
  \left| \Phi^{1}_{x}\right\rangle
  =N_{1}(|2,0;0,2\rangle+|0,2;2,0\rangle - c_{1}|1,1;1,1\rangle).
\end{equation}
Alternatively, for $x < \alpha(1+\text{cos}\theta)$, we get the asymmetric state
\begin{equation}\label{}
  \left| \Phi^{0}_{x}\right\rangle
  =(|2,2;0,0\rangle+|0,0;2,2\rangle)/\sqrt{2},
\end{equation}
up to a phase shift $\varphi_x = -\alpha\sin \theta ({x - 2\alpha \cos \theta})/2 \bmod 2\pi$ on the spatial mode $b_2$ according to the value of the measurement.

That is, if one permits the phase shift with feedback from the value of the measurement, then the twin-beam four-photon asymmetric state $\left| \Phi^{0}_{x}\right\rangle$ can be prepared.
Also, for the symmetric state, it is interesting to note that the output state is similar to the input state, up to a correlation coefficient.
Especially, for $c=1$, i.e.
\begin{equation}\label{}
  \left| \Phi\right\rangle=(|2,0;0,2\rangle+|0,2;2,0\rangle-|1,1;1,1\rangle)/\sqrt{3},
\end{equation}
it exactly is the twin-beam entangled state emitted by the SPDC source with the second-order term, and we here refer to this pair of the indistinguishable four-photon entangled states as \emph{two-pair}, for simplicity.
Obviously, for this two-pair, one can immediately obtain the result that the output state is the same as the input.

\section{Two-pair entanglement purification by cascading symmetry detectors}

\begin{figure}
  \includegraphics[width=5.2in]{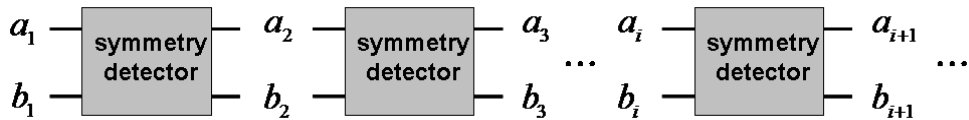}\\
  \caption{Efficient optical quantum circuit for purifying two-pair by cascading symmetry detectors.}
  \label{}
\end{figure}

As an important application of such symmetry detectors, we now present a scalable scheme for purifying the two-pair from the mentioned four-photon entangled states.
For this purpose, we construct a quantum circuit diagram by cascading these symmetry detectors, as shown in Fig.2, where the input/output modes correspond to the signal photons.
In each symmetry detector, we here simplify the initial model straightforwardly by discarding the result of the asymmetric state.

Then, after the $i$th cascading, one can obtain the symmetric state
\begin{equation}\label{}
  \left| \Phi^{i}_{x}\right\rangle
  =N_{i}(|2,0;0,2\rangle+|0,2;2,0\rangle - c_{i}|1,1;1,1\rangle),
\end{equation}
where $c_{i}=2/(1+c_{i-1})$, $N_{i}^2=1/(2+c_{i}^2)$.
The total success probability reads
\begin{equation}\label{}
 P=\prod_{i}P_{i}, ~~~ P_{i}=1/\{1+(1-c_{i-1})^2/[2+(1+c_{i-1})^2]\}.
\end{equation}
As a result, it is not difficult to find that such a cascading symmetry-detector is capable of purifying two-pair from the four-photon entangled states (\ref{4-photon-state-c}).

Clearly, we here take $c=2$ for example.
We calculate the iterative coefficient $c_i$ and the probability $P_i$ and plot the relationships of the correlation coefficients and the success probabilities versus the number of iterations (10 times), as shown in Fig.3.
The result shows that with a few iterations the correlation coefficient approaches $1$ and the success probability gets close to $1$.
Furthermore, since we take only those events into account that yield the required results via postselection, the present scheme of entanglement purification is near deterministic.

\begin{figure}
\centering
 \subfigure[]{
    \label{c-0-10} 
    \includegraphics[width=0.45\textwidth]{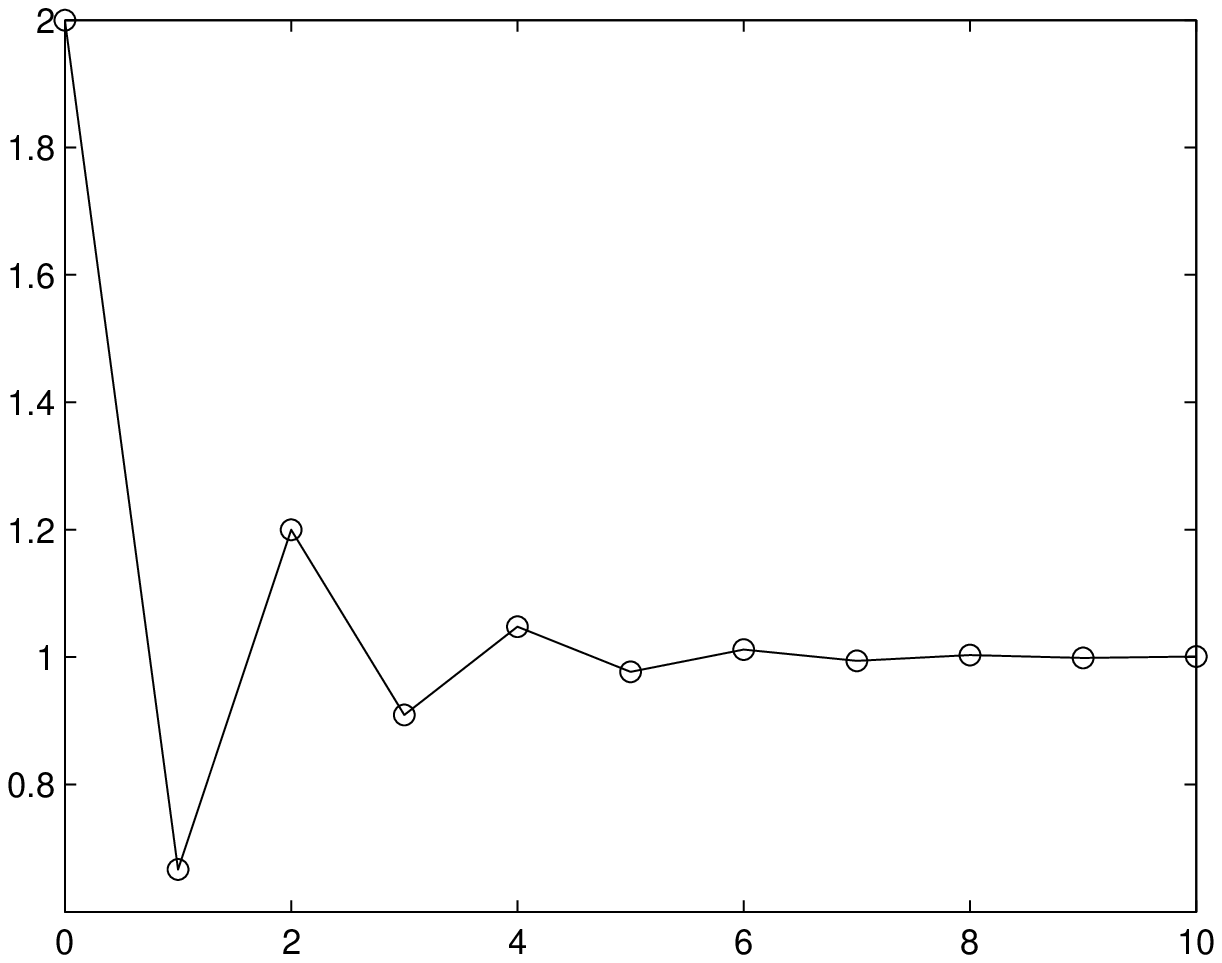}}
  \hspace{0.5in}
  \subfigure[]{
    \label{p-10} 
   \includegraphics[width=0.45\textwidth]{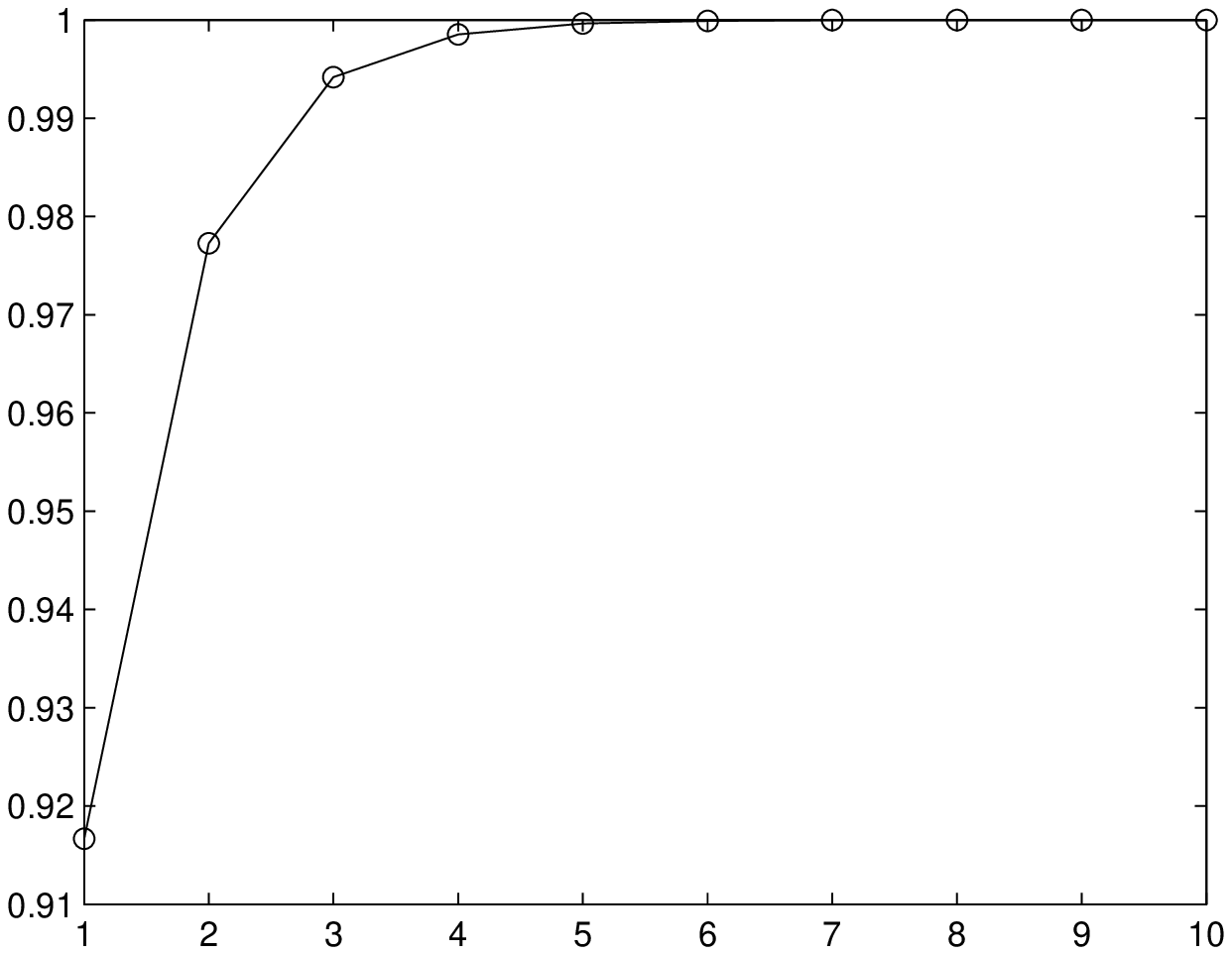}}
 \caption{Two relationships of the coefficients $c_{i}$ and probabilities $P_{i}$ versus the number of iterations $i$ ($i=1,2,\cdots,10$). By a few iterations, (a) $c_{i}$ gets close to $1$ and (b) $P_{i}$ may quickly be close to $1$.}
\end{figure}

\section{Symmetry detector for the higher-order emissions}

So far, we have addressed the symmetry detector and its interesting application, involving the second-order emission of the SPDC process.
In order to enable us to explore  multiphoton entanglement with a large number of photons from SPDC source,  we now describe a method to generalize the symmetry detector from the second-order to the higher-order emissions of the SPDC source.

For the higher-order emissions, consider an $n$-pair multiphoton entangled state $|\psi_n^{-}\rangle$.
When the photons passing through the $50:50$ BS, the transformation between the incoming modes ($a_1$ and $b_1$) and the outgoing modes ($a_2$ and $b_2$) is
\begin{eqnarray}\label{}
({a_{1_H}^{\dag}}{b_{1_V}^{\dag}} - {a_{1_V}^{\dag}}{b_{1_H}^{\dag}})^{n}
\rightarrow
({a_{2_H}^{\dag}}{b_{2_V}^{\dag}} - {a_{2_V}^{\dag}}{b_{2_H}^{\dag}})^{n}.
\end{eqnarray}
Then, as the multiphoton interference effect at the symmetric BS, the input $n$-pair entangled state $|\psi_n^{-}\rangle$ will be transformed into
\begin{eqnarray}\label{}
|\Phi_\text{BS}^n\rangle &=& \frac{1}{\sqrt{n+1}}\frac{1}{n!}({a_{2_H}^{\dag}}{b_{2_V}^{\dag}} - {a_{2_V}^{\dag}}{b_{2_H}^{\dag}})^{n}|0\rangle
\nonumber \\
&=&\frac{1}{\sqrt{n+1}} \sum_{k=0}^{n}(-1)^k |n-k\rangle_{a_{2_H}}|k\rangle_{a_{2_V}}|k\rangle_{b_{2_H}}|n-k\rangle_{b_{2_V}}.
\end{eqnarray}
This result implies that when the photons passed through a symmetric BS the $n$-pair entangled state remains unchanged.

In the process of the nonlinear interactions, for clearer statement, we here rewrite the Kerr phase shifts $\theta/3$ and $2\theta/3$ in spatial modes $a_2$ and $b_2$, and the original phase shift $-5\theta$ is accordingly replaced by $-\theta$.
On the basis of the methods of exploring multiphoton entanglement via weak  nonlinearities \cite{DYG2014, HDYG2015OE},
for arbitrary $m$-pair, $1 \leq m \leq n$, the total phase shift in the probe mode is $(m-1)\theta$.
Then after an $X$ homodyne measurement, one may obtain the $n$-pair with the value $x < \alpha \{ {\cos [(n-1)\theta]  + \cos [(n-2)\theta]} \}$,
$(n-1)$-pair with $\alpha \{ {\cos [(n-1)\theta]  + \cos [(n-2)\theta]} \} < x < \alpha \{ {\cos [(n-2)\theta]  + \cos [(n-3)\theta]} \}$, $\cdots$,
$m$-pair with the value $\alpha \{ {\cos (m\theta)  + \cos [(m-1)\theta]} \} < x < \alpha \{ {\cos [(m-1)\theta]  + \cos [(m-2)\theta]} \}$, $\cdots$,
two-pair with $ \alpha [\cos\theta+\cos2\theta] < x < \alpha [1+\cos\theta] $
or one-pair (the singlet state) with $x > \alpha [1+\cos\theta]$.
Obviously, as a particular case of the $n$-pair emissions, a one-pair emission is simple but instructive.
Since these multipair structures are robust against losing of photons they are maybe contribute to explore multiphoton entanglement from microscopic to macroscopic systems.

\section{discussion and summary}

By now, we have concentrated on the means to explore symmetry detectors for twin-beam multiphoton entanglement.
A realistic SPDC source with the higher-order emissions, however, inevitably emits one-pair, two-pair or $n$-pair entangled photons, spontaneously.
Another application of the present symmetry detector is thus to determine the number of the pairs emitted indistinguishably from the SPDC source, and then collect them.
Indeed, after the $X$ homodyne measurement on the probe beam one can immediately infer the number of the pairs by means of the value of measurement.
Also, the signal photons are specifically projected onto a particular multipair entangled state.

In conclusion, we explore an efficient symmetry detector for detecting the twin-beam multiphoton entanglement based on BS and weak nonlinearities.
Especially, as a typical application we suggest a scalable scheme of two-pair entanglement purification from a class of four-photon entangled states by cascading these symmetry detectors. Note that such two-pair entangled states may be very useful for multiphoton quantum information processing in the future.
In the present architectures, there are several remarkable advantages.
First, for the higher-order emissions, it is capable of determining the number of the pairs emitted indistinguishably from the SPDC source.
Second, since we here only use a symmetric BS and two small Kerr nonlinearities, our symmetry detector is simple and novel.
At last, it is possible to extend our means to general circuits constructed from linear elements, SPDC sources, and detectors.
We hope that our scheme will stimulate investigations on the applications of higher-order emissions from the SPDC source.

\section{Acknowledgements}
This work was supported by the National Natural Science Foundation of China under Grant Nos: 11475054, 11371005, 11547169, Hebei Natural Science Foundation of China under Grant No: A2014205060, the Research Project of Science and Technology in Higher Education of Hebei Province of China under Grant No: Z2015188, Foundation for High-level Talents of Chengde Medical University under Grant No: 201701, Langfang Key Technology Research and Development Program of China under Grant No: 2014011002.

\end{document}